# A Model-Driven Architecture Approach for Developing Healthcare ERP: Case study in Morocco

Fatima Zahra Yamani[1], Mohamed El Merouani[2]

[1] Department of Science and Technology, Center of Doctoral Studies, Abdelmalek Essaadi University
Tetouan, 93000, Morocco

[2] Department of Science and Technology, Center of Doctoral Studies, Abdelmalek Essaadi University
Tetouan, 93000, Morocco

**Abstract**
Nowadays, there are many problems in the Enterprise Resource Planning (ERP) implemented in the majority of hospitals in Morocco such as the difficulty of adaptation by the different users, the lack of several functionalities, errors that block the daily work, etc. All these problems require frequent modifications in the code, which implies a high effort to develop healthcare ERP as one of complex systems. In this paper, we are going to present a model-driven approach for developing healthcare ERP based on class diagram. First, we constitute the independent model using UML, define the transformation rules then apply them on our source model class to generate at the end an XML file that will be necessary for the ERP code. Our approach will not only resolve the above problems, but also improve the efficiency of software development through the automatically generated code.
*Keywords:* Healthcare ERP, UML, Model Driven Architecture (MDA), transformation by modeling, PIM, MOF 2.0 QVT.

## 1. Introduction

For many years, the healthcare sector was little concerned by Information Technology (IT), while in the industry; IT was a major challenge for the development and durability of companies.

Currently, no hospital can continue using old practices and technologies. It is imperative that hospitals adopt the latest trends in technology and insight to retain their users, for this, there is strong pressure to develop ERP (Enterprise Resource Planning) in hospitals in order to make them more efficient and able to meet the expectations of patients and the need of users.

Despite all attempts to develop healthcare ERP in Morocco, there are still many problems during development. In fact, user requirements are always changing, which implies frequent modifications on models, it is necessary to modify separately and manually the analysis model, design model and the code. All of these tasks result in increased maintenance and time costs as well as inconsistency between requirements definition, analysis, design and implementation.

In this context, the Object Management Group (OMG) proposed Model Driven Architecture (MDA) in 2001 as an approach to design, implement and develop complex applications with a great initial effort to specify the features.

The principle key of MDA is to rely on the Unified Modeling Language (UML) standard to describe models separately at different phases of the application development cycle. MDA aims to highlight the intrinsic qualities of models, such as durability, productivity and taking into account the execution platforms.

Due to the lack of research in this sector in Morocco, further studies should discuss and investigate the development of ERP in Moroccan healthcare organizations. Our goal will be then identifying the improvement perspectives of healthcare ERP in order to have an easy-to-use one, covering all areas of health activity, having modules adapted to all types of hospital organizations in Morocco, as well as proposing a MDA approach for ERP healthcare development. This approach includes UML modeling and its transformation to generate a source code.

After this introduction, this document is divided as follows. Section 2 presents the most relevant related works. Section 3 defines the MDA approach while the section 4 describes the transformations between models. The MDA approach for Healthcare ERP is the main topic of section 5. In section 6, we present the result of code generation process. Finally, section 7 sums up the main conclusion.

## 2. Related Works

In the last few years, different studies in several disciplines have been interested in the MDA approach; it has become the subject of great interest for many research teams (E-learning [1, 3], Mobile Application [2], Web-Marketing [4], etc.)

The author in [1] relied on the MDA approach to have a multi-target learning management system generator. He intended to simplify the design and the development of e-learning platforms, while the approach presented in [2] is used to model and generate mobile applications. The approach includes UML modeling and automatic code generation using Acceleo. They could, in the end, develop all the necessary meta-classes to generate a mobile application then use Acceleo as a transformation language.

Xiao Cong and the other authors in [3] have proposed a model-driven development approach for E-Learning platforms, after a business logic analysis, they establish a CIM model, stratified on the PIM under the J2EE framework and proposed the method of transformation from PIM to PSM.

In [4] authors have applied the MDA approach to generate the N-tiers web application based on UML class diagram, creating a skeleton of a social network. As a transformation language, they used the MOF 2.0 QVT (Meta-Object Facility 2.0 Query-View Transformation) standard to define the meta-model for the development of model transformation.

The authors of the work [5] propose an approach for transforming a CIM into a PIM using the core modeling concepts of the UML. They have described some important cases of transformation from CIM to PIM and propose a new approach by modeling to realize the transformation.

This paper aims to benefit from the experience of other authors in their application of MDA and apply it in the sector that is relevant to the healthcare sector. Actually, it is the only known work aimed at reaching this goal in Morocco.

## 3. Model Driven Architecture (MDA)

The Model driven architecture (MDA) is an approach proposed in late 2000 by the OMG, a consortium of over 1 000 companies. It is an approach to software design, development and implementation [6].

The MDA approach advocates the massive use of models at the different phases of application development to generate automatically a code source and offers first responses to how, when, what and why to model. It aims to highlight the intrinsic qualities of models, such as durability, productivity and taking into account the execution platforms.

Among the main terms defined by the OMG, we have: model, meta-model and meta-metamodel, defined as follows:

- Model: is an abstraction, a simplification of a system that helps to understand the modeled system and to answer the questions asked about it [3]. The model layer is comprised of the meta-model that describes data in the information layer.
- Meta-model: is a model of a model, is a model that defines the concepts and rules of a modeling language. It defines the object types that can be used to represent a model, relations between object types, attributes of the object types and rules to combine object types and relations.
- Meta-metamodel: model defines the language in which a meta-model can be expressed.

MDA relies also on the UML standard as a principle key to describe separately models at different phases of an application's development cycle, it defines three levels of models advocated for the construction of software: Computation Independent Model (CIM), Platform Independent Model (PIM), Platform Specific Model (PSM) and code, presented in Figure 1 and defined just after.

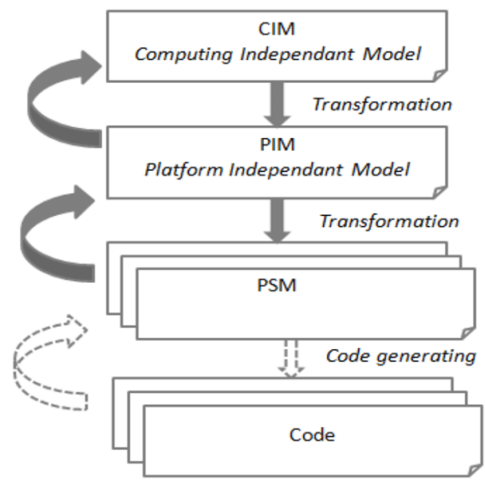

Fig. 1 Model Driven Architecture levels

- CIM: The objective is to create a requirements model for the future application. Such a template should represent the application in its environment to define what services are offered by the application. It is important to note that requirements model does not contain information about the implementation of the application or the treatments. With UML, requirements model can be presented as a use case diagram.
- PIM: Once the requirements model is completed, the analysis and design work can begin. In the MDA approach, this phase also uses models. The role of analysis and design models is to be sustainable and to make the link between the requirements model and the application code [7]. These models must also be productive since they form the basis of the entire code generation process defined by MDA.
- PSM: Is the model that comes closest to the final code of the application [7]. A PSM is a code model that describes the implementation of an application on a particular platform, so it is linked to an execution platform.

Other than the UML standard, MDA includes the definition of other standards as MOF and XMI.

## 4. Model To Model Transformation (M2M)

The MDA provides a process of converting a model into another model of the same system, those model transformations are an essential part of model-driven engineering approach to software development. The main transformations recommended by MDA are: CIM transformation to PIM and PIM transformation to PSM. In our paper, we are interested in the second transformation PIM to PSM; this will be applied on our healthcare ERP model.

Among the fundamental approaches of model transformations, there is: approach by Modeling, approach by Template and approach by Programming, in the present work we chose approach by Modeling. This approach resides of applying concepts from model engineering to model transformations. The objective is modeling a transformation, to reach perennial and productive transformation models and to express their independence towards the platforms of execution [4].

Figure 2 presents the approach by modeling. According to MOF 2.0 QVT, a standard transformation language elaborated by the OMG, the transformation of model is defined as a structured model. In the transformation, the MOF 2.0 QVT defines the rules between the source and target meta-model. This model needs to be transformed to execute the transformation on an execution platform.

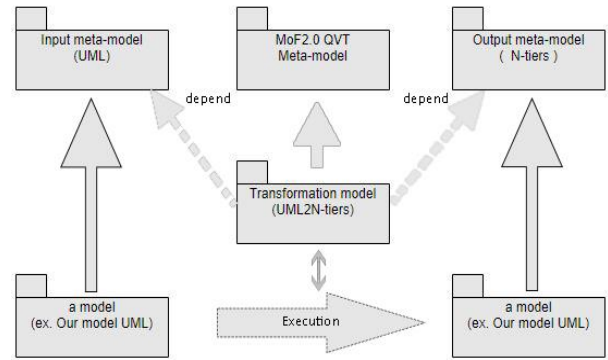

Fig. 2 Approach by modeling

### 4.1. MOF 2.0 QVT

Using the modeling approach is designed to have a productive models' transformation, independently of any execution platform. For this reason, the OMG issued on April 2002 a request for proposals for a standard to this transformation language, which is the MOF (Meta-Object Facility) 2.0 QVT (Query, Views, and Transformations), terms are defined as follows:
- Query: is an expression that is evaluated over a model. The result of a query is one or more instances of types defined in the source model, or defined by the query language [8].
- View: A view is a model that is completely derived from another model, it represents the user interface.
- Transformation: A transformation generates a target model from a source model.

This standard defines the meta-model for the development of transformation model. The standard QVT is built in a modular way and it combines several paradigms. Through hybrid architecture, it offers the combined benefits of declarative approaches and imperative approaches. Basically, the imperative approach is better adapted for complex transformations including a significant algorithm component while the declarative approach; it has the advantage of optional case management in a transformation.

In this paper, we are going to use the imperative approach. The imperative QVT component is supported by Operational Mappings language (OML), one of the three languages included by the MOF QVT specification. Currently, Operational Mappings is the best supported variant in terms of tools. SmartQVT is the first open source implementation of the QVT Operational Mapping language. This tool is developed by France Telecom. It is an Eclipse plugin under EPL license running on top of

EMF framework. It is composed of three components, which are:
- The code editor: this component helps the user to write QVT code by highlighting keywords.
- The parser: this component converts QVT code files into model representations of the QVT programs.
- The compiler: this component converts model representations of the QVT program into executable Java programs.

## 4.2. PIM to PSM transformation

### 4.2.1. PIM source meta-model

Following the definition presented in section 3, the PIM is also known as the analysis and design model, is an abstract model independent from any running platform. It ignores operating system, programming languages, hardware and networking. It is designed to describe the know-how or business knowledge of an organization.

As it is illustrated in Figure 3, the source meta-model form a simplified UML model (class diagram) based on packages containing data types and classes. The classes are composed of structural features represented by attributes and behavioral characteristics represented by operations.

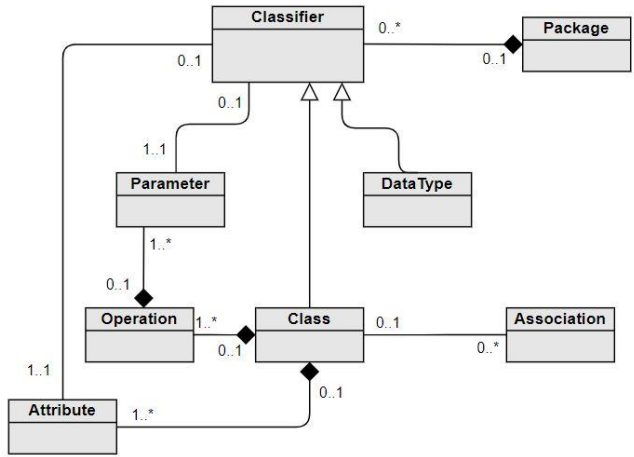

Fig. 3 Simplified UML Metamodel

- Package: A technique that allows implementing partitioning models while preserving the consistency of the whole. A package is a collection of modeling elements: classes, associations, objects, components, packages and models.
- Classifier: is an abstract meta-class classification concept that serves as a mechanism to show interfaces, classes, datatypes and components. It describes a set of instances that have common behavioral and structural features.
- DataType: Data types are model elements that define data values.
- Class: A class is the formal description of a set of objects with semantics and common characteristics.
- Parameter: expresses the concept of parameters of an operation. It explains the link between Parameter meta-class and Classifier meta-class.
- Operation: Functionality ensured by a class. The description of an operation can specify the input and the output parameters and the elementary actions to be performed.
- Attribute: is type of basic information that is part of the structure of a class (especially an entity).

### 4.2.2. PSM target meta-model

In this section we present the various meta-classes forming the meta-model target, it is manly composed of three essential parts. The first part of the target meta-model (illustrated in Figure 4) presents the different meta-classes to express the concept of DAO contained in the DaoPackage. The meta-model is composed of:
- **Interface**;
- **Table**;
- **DaoPackage** that represents the package containing the meta-classes to express the concept of DAO;
- The **HibernateDaoSupport** presenting the concept of generic class for DAOs;
- **IDao**, which expresses the concept of Dao interface that contain the definition of methods to create, remove, update, and display;
- **Pojo** expresses the objects that will communicate with the tables of relational database;
- **DaoImpl** presents the concept of Dao implementation, it contains methods to create, remove, update, and display data in the database; and finally,
- The **CrudProjectPackage** that is connected to the meta-class **DaoPackage**.

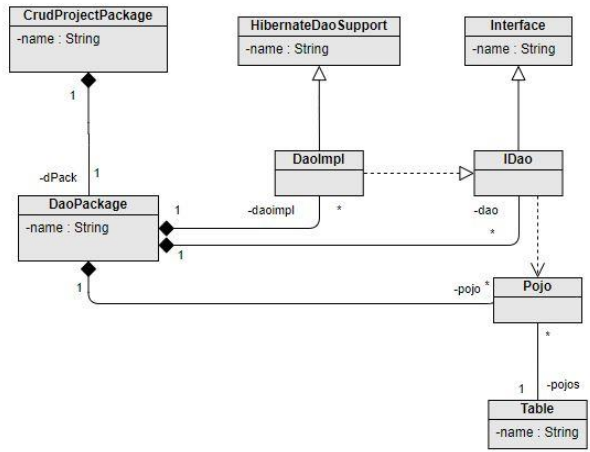

Fig. 4 Simplified DaoPackage meta-model

The business model is the second part of the target meta-model, in figure 5, we present the different meta-classes expressing the concept of DI contained in the Business Package. The components shown in the figure are:

**BusinessPackage** defining the package and containing the different meta-classes to show the concept of the business logic of target application; **IService** expressing the concept of service interface that contains the definition of methods; the methods representing in IDao meta-class and declared in IService meta-class are defined in **ServiceImpl**; and finally, **Dto** represents the concept of business object, without forgetting the **IDao** and **Pojo** that are already defined in the DaoPackage meta-model.

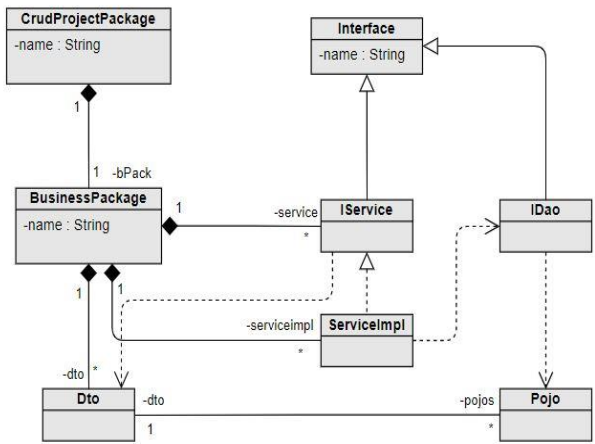

Fig. 5 Simplified BusinessPackage meta-model

The third part of the target meta-model is presented in figure 6 as UIPackage, it represents a concept of MVC2 implementation in the user interface. The component of this meta-model is a **UIPackage** connected to the meta-class ViewPackage and ControllePackage representing View and Controller package, **DelagatingActionProxy** that defines the concept of Proxy for a Spring-managed Struts Action that is presented in WebApplicationContext. The proxy is defined in the Struts config file, specifying this class as action class. It will delegate to a Struts Action bean in the ContextLoaderPlugIn context [9], **ActionMapping** that represents the concept of ActionMapping classes, **Action** representing the class containing its own processing of the application [10], **ActionForm** represents a form that contain the parameters of the request from the view, **JspPage** represents a Jsp page, through a hyperlink in a Jsp, an action class may be called, **HttpRequest** is the concept of HttpServletRequest classes, **HttpResponse** expresses the concept of HttpServletResponse classes, **ApplicationContext** defines the concept of Central interface to afford configuration for an application and finally **ServiceLocator** that represents the concept of Service lookup and creation involves complex interfaces and network operations.

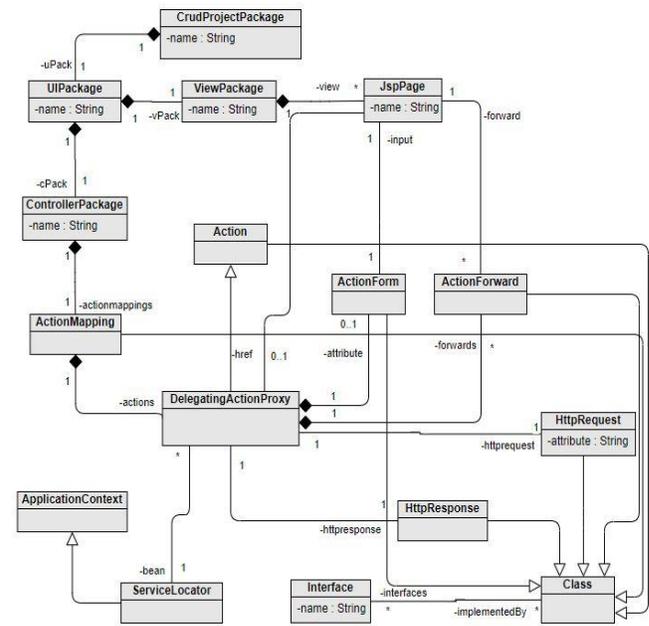

Fig. 6 Simplified UIPackage meta-model

4.2.3. Transformation rules

*Main algorithm:*
*input umlModel:UmlPackage*
*output crudModel:CrudProjectPackage*
*begin*
*create CrudProjectPackage crud*
*create DaoPackage daoPackage*
*for each e ∈ source model*

```
x = transformationRuleOnePojo(e)
link x to dp
x = transformationRuleOneIDao(e)
link x to dp
x = transformationRuleOneDaoImpl(e)
link x dp
end for
create BusinessPackage bp;
for each pojo ∈ target model
x = transformationRuleTwoDto(pojo)
link x to bp
end for
for each e ∈ source model
x = transformationRuleTwoIService(e)
link x to bp
x = transformationRuleTwoSrviceImpl(e)
link x to bp
end for
create UIPackage uip;
create ViewPackage vp
vp = transformationRuleThreeView(e)
create ControllerPackage cp
cp = transformationRuleThreeController(e)
link vp to uip
link cp to uip
link dp to crud
link bp to crud
link uip to crud
return crud
end
function
transformationRuleOnePojo(e:Class):Pojo
begin
create Pojo pj
pj.name = e.name
pj.attributes = e.properties
return pj
end
function
transformationRuleOneIDao(e:Class):IDao
begin
create IDao idao
idao.name = 'I'+e.name+ 'Dao'
idao.methods = declaration of e.methods
return idao
end
function
transformationRuleOneDaoImpl(e:Class):DaoImpl
begin
create DaoImpl daoImpl
daoImpl.name = e.name+ 'DaoImpl'
for each e1 ∈ DaoPackage
if e1.name = 'I'+e.name+ 'Dao'
put e1 in interfaces
end if
end for
link interfaces to daoImpl
return daoImpl
end
function

transformationRuleTwoDto(p:pojo):Dto
begin
create Dto dto
dto.name = p.name
dto.attributes = p.attributes
return dto
end
function
transformationRuleTwoIService(e:Class):IService
begin
create IService iservice
iservice.name = 'I'+e.name+ 'Service'
iservice.methods = declaration of e.methods
return iservice
end
function
transformationRuleTwoServiceImpl(e:Class):Service Impl
begin
create ServiceImpl serviceImpl
serviceImpl.name = e.name+ 'ServiceImpl'
for each e1 ∈ BusinessPackage
if e1.name = 'I'+e.name+ 'Service'
put e1 in interfaces
end if
end for
link interfaces to ServiceImpl
return ServiceImpl
end
function
transformationRuleThreeView(e:Class):ViewPackage
begin
create ViewPackage vp
for each e ∈ source model
if e.methods.name ≠ 'remove'
create JspPage page
link page to vp
end if
end for
return vp
end
function
transformationRuleThreeController(e:Class):ControllerPackage
begin
create ControllerPackage cp
create ActionMapping am
for each page viewPackage
link page to actionForward
create actionForm
create Action action
create ActionForward actionForward
actionForm.input=page
actionForm.attribute=action
link page to actionForward
link actionForward to action
put action in am
end for
link am to cp
return cp
end
```

The algorithm used for the transformation of UML model source to N-tiers target model is written with QVT language. As we can observe, the method of the entry makes the correspondence between the elements of the UMLPackage type of the input model and the element of the CrudProjectPackage type of the output model. With the creation of the elements of type package 'Dao', 'Business' and 'Presentation, the transformation of the UML package into N-tiers package will be done. Each class of package UML will be transformed to Jsp page and Action in the View package, to DTO, IService and ServiceImpl in the Business package, and to Pojo, IDao and DaoImpl in the Dao package.

## 5. Healthcare ERP: Case study

The healthcare organizations need to have automated information systems, such as ERP in order to meet the quality requirements of health services. It helps to streamline the processes of the healthcare organization, to manage and control various departments. The healthcare ERP provides complete solutions to different segments of the health industry and solves all the problems of doctors, nurses, pharmacists, etc.

The main features we want to model in healthcare ERP are N-Tier Architecture and streamlines healthcare processes.

### 5.1. Building PIM of ERP Healthcare

The software is divided into different modules, each one dedicated to a specific activity of the hospital. In this paper, we present only a part of "laboratory module", that is considered one of the important modules in any hospital. Figure 7 illustrates an instance of UML Model. The PIM constructed respect the elements on the above UML meta-model.

Our case study adopts the CRUD operations (Create, Remove, Update, and Display) that are often implemented in all systems.

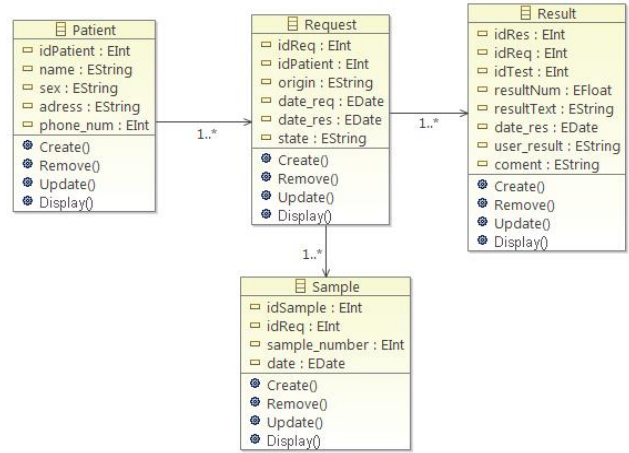

Fig. 7 Instance of UML Model.

### 5.2. Generating the PSM

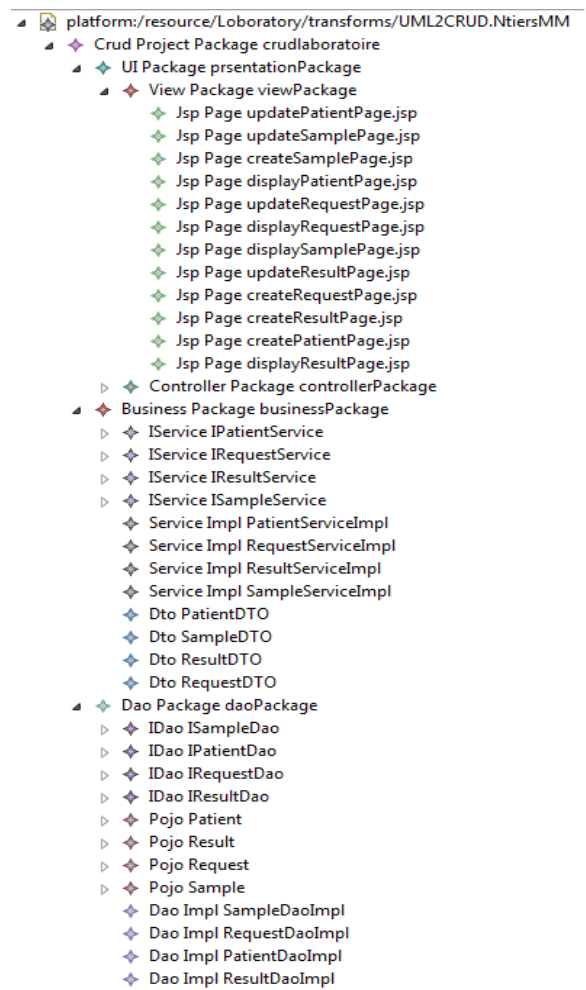

Fig. 8 Generated PSM Laboratory Model

As it is shown in figure 8, the UIPackage is at the head of the elements of the generated PSM model, it contains viewPackage contaning the JSPs, that is DisplayPatientPage.jsp, DisplayRequestPage.jsp, DisplayResultPage.jsp, DisplaySamplePage.jsp, CreatePatientPage.jsp, CreateRequestPage.jsp, CreateResultPage.jsp, CreateSamplePage.jsp, UpdatePatientPage.jsp, UpdateRequestPage.jsp. UpdateResultPage.jsp, UpdateSamplePage.jsp. We can note in the figure that the Remove operation does not exist with other CRUD operations, this is because it requires any form. Just after, we go to the controllerPackage element that contains ActionMapping, this ActionMapping contains twenty four delegating action proxy whose names are respectively DisplayXAction, CreateXAction, UpdateXAction, RemoveXAction, CreateXEndAction, UpdateXEndAction, where X should be replaced by Patient, Request, Result and Sample. The next element in the generated PSM model is businessPackage, it contains four services interfaces, four services implementations and four Dtos for each of the objects "Patient", "Request", "Result" and "Simple". All of the Pojos object, the four Daos' interfaces that contains methods and Daos implementations are part of the last element in the generated PSM model.

## 6. Results of code generation process

We present in this section the result of code generation process. In fact, we modeled and generated the code of the whole module, but we display only the part that includes classes shown above. It introduces in Figure 9 DataTypes, classes, operations and properties.

```xml
<?xml version="1.0" encoding="UTF-8"?>
<NtiersMM:CrudProjectPackage xmi:version="2.0" xmlns:xmi="http://www.omg.org/XMI"
xmlns:NtiersMM="http:///NtiersMM.ecore" name="crudlaboratoire">
  <uPack name="prsentationPackage">
    <vPack name="viewPackage">
    <cPack name="controllerPackage">
  </uPack>
  <bPack name="businessPackage">
    <services name="IPatientService" implementedBy="//@bPack/@serviceimpl.0">
    <services name="IRequestService" implementedBy="//@bPack/@serviceimpl.1">
    <services name="IResultService" implementedBy="//@bPack/@serviceimpl.3">
    <services name="ISampleService" implementedBy="//@bPack/@serviceimpl.2">
    <serviceimpl name="PatientServiceImpl" interfaces="//@bPack/@services.0"/>
    <serviceimpl name="RequestServiceImpl" interfaces="//@bPack/@services.1"/>
    <serviceimpl name="SampletServiceImpl" interfaces="//@bPack/@services.3"/>
    <serviceimpl name="ResultServiceImpl" interfaces="//@bPack/@services.2"/>
    <dto name="PatientDTO" pojos="//@dPack/@pojo.1"/>
    <dto name="SampletDTO" pojos="//@dPack/@pojo.0"/>
    <dto name="ResultDTO" pojos="//@dPack/@pojo.2"/>
    <dto name="RequestDTO" pojos="//@dPack/@pojo.3"/>
  </bPack>
  <dPack name="daoPackage">
    <dao name="ISampleDao" implementedBy="//@dPack/@daoimpl.0">
    <dao name="IPatientDao" implementedBy="//@dPack/@daoimpl.3">
    <dao name="IRequestDao" implementedBy="//@dPack/@daoimpl.2">
    <dao name="IResultDao" implementedBy="//@dPack/@daoimpl.1">
    <pojo name="Sample" dto="//@bPack/@dto.1">
    <pojo name="Patient" dto="//@bPack/@dto.0">
    <pojo name="Result" dto="//@bPack/@dto.3">
    <pojo name="Request" dto="//@bPack/@dto.2">
    <daoimpl name="SampleDaoImpl" interfaces="//@dPack/@dao.1"/>
    <daoimpl name="RequestDaoImpl" interfaces="//@dPack/@dao.3"/>
    <daoimpl name="PatientDaoImpl" interfaces="//@dPack/@dao.2"/>
    <daoimpl name="ResultDaoImpl" interfaces="//@dPack/@dao.0"/>
  </dPack>
</NtiersMM:CrudProjectPackage>
```

Fig. 9 XML file generated

## 7. Conclusion

In this paper, we proposed a model driven approach to develop a Healthcare ERP. Unlike the traditional software development process, model transformation in our approach ensures consistency by defining model transformation rules and resolves problems of difficulty in ensuring consistency in requirements definition, analysis and design. For this, we had a comprehensive analysis of the problems in the development of healthcare ERP and we focused on the generation phase based on MDA approach.

The goal of our approach is to develop all meta-classes needed to generate N-tiers application, we applied the approach by modeling and we used the UML class diagram as a model of PIM, and then used the MOF 2.0 QVT standard as a transformation language.

Through the transformation rules defined, an XML file has been generated, based on the source model instance class diagram presented, that file can be used to produce the necessary code of the target application.

**Fatima Zahra Yamani** got her Master Degree in IT management from Abdelmalek Essaadi University of Tetouan, Morocco in 2016. She is currently Ph.D student in Science and Technology with the research team of Modeling and Computer Theory at the center of doctoral studies in Tetouan, Morocco. Her research activities have focused on healthcare ERPs in Morocco using Model Driven Architecture approach.

**Mohamed El Merouani** obtained his Ph.D in Mathematics from University of Granada, Faculty of Sciences. He was a responsible of Department of Statistics and IT at the polydisciplinary faculty of Tetouan during many years. He is a professor and responsible of "Modeling and Computer Theory" research team. Currently, he is professor at the University Abdelmalek Essaadi of Tetouan (Morocco). He teaches several courses in the domain of mathematics and statistics.